\documentclass{article}
\usepackage{amsmath, amsfonts, amssymb}
\usepackage{authblk}
\usepackage[super, sort&compress, comma]{natbib}
\usepackage{graphicx,color}
\usepackage[colorlinks = true, citecolor = blue, linkcolor = red]{hyperref}
\usepackage{xspace}
\usepackage{upgreek}

\newcommand{\x}{\ensuremath{\underline{r}}}
\newcommand{\nm}{\ensuremath{\,\mathrm{nm}}\xspace}
\newcommand{\nA}{\ensuremath{\,\mathrm{nA}}\xspace}
\newcommand{\pA}{\ensuremath{\,\mathrm{pA}}\xspace}
\newcommand{\T}{\ensuremath{\,\mathrm{T}}\xspace}
\newcommand{\V}{\ensuremath{\,\mathrm{V}}\xspace}
\newcommand{\mV}{\ensuremath{\,\mathrm{mV}}\xspace}

\newcommand{\meV}{\ensuremath{\,\mathrm{meV}}\xspace}

\newcommand{\ECn}{\ensuremath{E_\mathrm{C}^n}\xspace}

\newcommand{\epsij}{\ensuremath{\varepsilon_j}\xspace}

\author[1]{Nils~M.~Freitag}
\author[2]{Tobias~Reisch}
\author[2]{Larisa~A.~Chizhova}
\author[1,3]{P\'eter~Nemes-Incze}
\author[1]{Christian~Holl}
\author[4]{Colin~R.~Woods}
\author[4]{Roman~V.~Gorbachev}
\author[4]{Yang~Cao}
\author[4]{Andre~K.~Geim}
\author[4]{Kostya~S.~Novoselov}
\author[2]{Joachim~Burgd\"orfer}
\author[2]{Florian~Libisch}
\author[1]{Markus~Morgenstern}

\affil[1]{II. Institute of Physics B and JARA-FIT, RWTH Aachen University, Otto-Blumenthal-Stra{\ss}e, 52074 Aachen, Germany}
\affil[2]{Institute for Theoretical Physics, TU Wien, Wiedner Hauptstra{\ss}e 8-10, 1040 Vienna, Austria}
\affil[3]{Centre for Energy Research, Institute of Technical Physics and Materials Science, Lend\"ulet - Topology in Nanomaterials Research Group, 1121 Budapest, Konkoly-Thege way 29-33, Hungary}
\affil[4]{School of Physics \& Astronomy, University of Manchester, Manchester M13 9PL, United Kingdom}

\title{Large tunable valley splitting in edge-free graphene quantum dots on boron nitride}

\begin{document}

\date{}

\maketitle

\textbf{
Coherent manipulation of binary degrees of freedom is at the heart of modern quantum technologies. Graphene offers two binary degrees: the electron spin and the valley.
Efficient spin control has been demonstrated in many solid state systems, while exploitation of the valley has only recently been started, yet without control on the single electron level.
Here, we show that van-der Waals stacking of graphene onto hexagonal boron nitride offers a natural platform for valley control.
We use a graphene quantum dot induced by the tip of a scanning tunneling microscope and demonstrate valley splitting that is tunable from -5 to +10 meV (including valley inversion) by sub-10-nm displacements of the quantum dot position.
This boosts the range of controlled valley splitting by about one order of magnitude.
The tunable inversion of spin and valley states should enable coherent superposition of these degrees of freedom as a first step towards graphene-based qubits.}
\\

Electrical control is a central requirement to exploit the binary degrees of freedom of a single electron in a scalable way \cite{Ladd2010}.
This has been realized for spin systems using, e.g., small shifts of the electron spin within the field of a micro-magnet \cite{PieroLadriere2008,WuX2014}.
The valley degree of electrons has recently been detected in transport experiments on graphene \cite{Abanin2011,YangC2013,Shimazaki2015,Sui2015,Ju2015,Wallbank2016}, but its control on the single electron level has not been achieved.
Alternative materials, such as Si \cite{Rahman2011}, offer only very small tuning ranges of the valley splitting by less than 0.5~meV \cite{YangC2013,Gokmen2008,Kobayashi2016,Gamble2016,Scarlino2017,Mi2017}.

The valley degree of freedom in graphene is a consequence of the honeycomb structure with its two atoms within the unit cell \cite{Xiao2007,Pesin2012}.
Hence, breaking the equivalence of the two atoms (sublattice symmetry breaking) is the natural avenue to break the valley degeneracy as a starting point for tuning. This indeed works straightforwardly, if the time reversal symmetry is additionally broken, e.g., by a magnetic field $B$ \cite{CastroNeto2009}.
The sublattice symmetry breaking can be achieved by van-der Waals stacking of 2D materials exploiting the different stacking of the two graphene atoms on top of the supporting atoms. This stacking moreover spatially varies due to the different lattice constants of the adjacent materials \cite{Geim2013,Woods2014,Novoselov2016}, implying a spatially varying valley splitting which we exploit in our experiment.

\begin{figure*}[h!]
    \includegraphics[width = \textwidth]{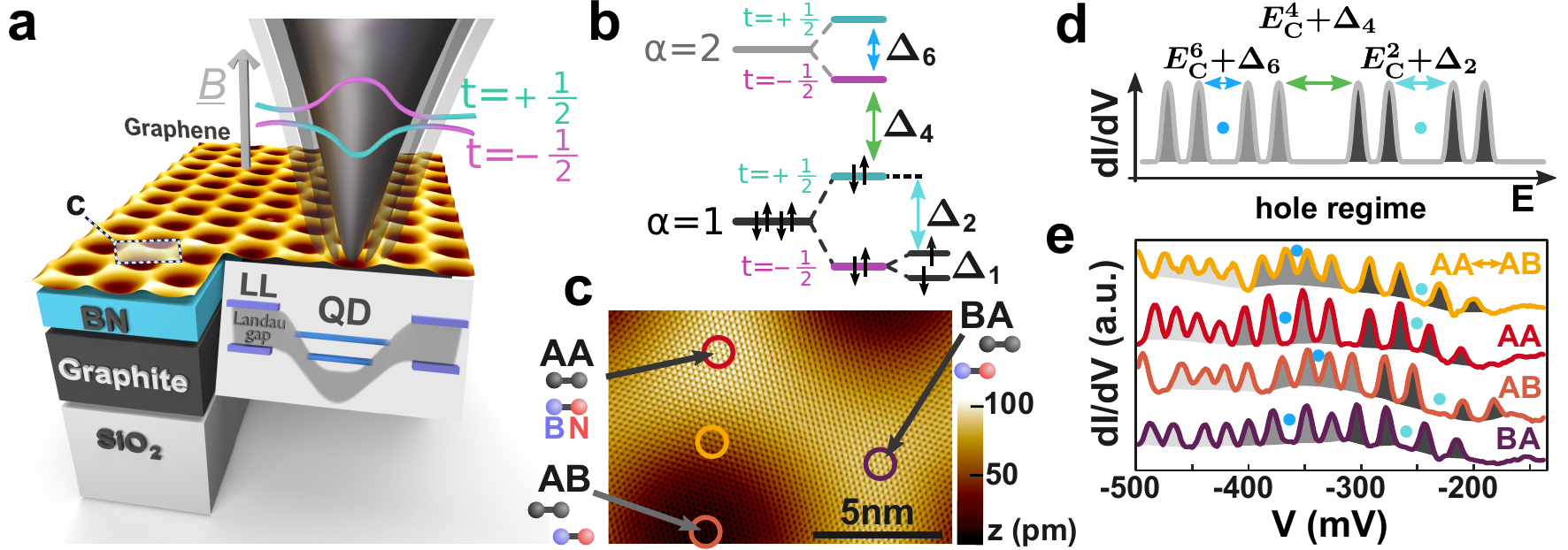}
    \caption{ \textbf{Edge free quantum dot.}
		\textbf{(a)} Sketch of the experiment: colored blocks on the left show the stacking sequence SiO$_2$/Graphite/hBN/Graphene \cite{Freitag2016}. The STM tip (grey cone) is moved above graphene deposited on BN with its honeycomb lattice collinearly aligned with that of BN  (brown-yellow STM image with BN-induced superstructure, $V = 300\mV$, $I = 1\nA$). A perpendicular $\underline{B}$ field (7~T, grey arrow) leads to Landau levels (LL, purple lines) and corresponding Landau gaps (grey area). The electric field of the tip induces band bending (curvature of Landau gap), leading to confined states (blue lines), hence, to a quantum dot (QD). The QD is moved by moving the STM tip above the superstructure (light grey areas around the cone). This modifies the confined state energies as the valley levels $\tau =1/2$ and $\tau = - 1/2$ associated with the $K$ and $K'$ points of the unperturbed band structure (cyan and magenta lines). The rectangle marked \textbf{c} indicates the area shown in magnification in \textbf{c}.
        \textbf{(b)} Schematic energy level diagram of the QD. The two orbital levels $\alpha=1$ and $\alpha=2$
exhibit valley splitting $E_{\alpha,\tau =+1/2,\sigma}-E_{\alpha,\tau = -1/2,\sigma}$. The Zeeman splitting $E_{\alpha,\tau,\sigma = +1/2}-E_{\alpha,\tau,\sigma= - 1/2}$ is small  ($\simeq 800\;\mu$eV) and only shown for the lowest valley state. The resulting energy distances $\Delta_n$ between adjacent levels are labeled with consecutive $n$. $\Delta_n$ for odd $n$ correspond to Zeeman splittings, which is only displayed for $n=1$.
		\textbf{(c)} Atomically resolved STM image of rectangular area marked in \textbf{a}, $V = 137\mV$, $I = 0.3\nA$. Different stacking areas (AA, AB, BA) are indicated by arrows with stick and ball models below the labels (C: gray, B: blue, N: red). Colored rings mark the positions of spectra in \textbf{e}.
		\textbf{(d)} Sketch of expected $dI/dV$ peak sequence for hole charging according to the level diagram in \textbf{b} using the same colored arrows and the same $\Delta_n$;
			\ECn: charging energy for filling of the $n^{\rm th}$ level. Blue dots highlight valley gaps.
		\textbf{(e)} $dI/dV$ spectra recorded at the positions encircled by the same color in \textbf{c} with corresponding stackings marked (AA$\leftrightarrow$AB: between AA and AB). Quadruplets of charging peaks, belonging to the same orbital, are shaded equally. Blue dots mark valley transitions. Predominant quadruplet sequences (yellow spectrum), predominant doublet sequences (purple spectrum), or a mixture of both (red and orange spectra) appear,			$V_\mathrm{stab}=1\V$, $I_\mathrm{stab}=700\pA$,
			$V_\mathrm{mod}=4.2\,\mathrm{mV}_\mathrm{rms}$, $B=7$~T, $T=8$~K.			
			}
		\label{fig:One}
\end{figure*}

We have recently demonstrated smoothly confined Dirac fermions in an edge-free graphene quantum dot (QD)
by combining the electric field of the tip with a perpendicular $B$ field (Fig.~\ref{fig:One}a) \cite{Freitag2016}.
The $B$ field quantizes the continuous spectrum of graphene in terms of Landau levels (LLs, LL spacing $\approx$ 100 meV at $B=7$ T) \cite{CastroNeto2009}.
The electric field of the tip exploits the energy gaps between LLs to
achieve edge-free confinement, i.e., it shifts energy levels from the LLs
into the gap \cite{Freitag2016}. We thereby overcome the well-known
problem of edge localization within etched graphene
QDs \cite{Bischoff2015}.
By confining without resorting to physical edges, these
dots preserve the two-fold valley and spin symmetries of pristine
graphene (Fig.~\ref{fig:One}b, d).

The charging of the confined
levels has been directly measured by tuning the voltage of the STM tip
such that the states cross the Fermi level $E_{\rm F}$.
This revealed
the most regularly spaced charging sequence of graphene QDs achieved
to date \cite{Freitag2016}.
The measured level separations have been reproduced with the help of tight binding (TB) calculations. Hence, the charging peaks
could be assigned to LLs and to particular orbital and valley states.
Most notably, we observe quadruplets of charging peaks belonging to a single
orbital quantum number of the dot and a partial splitting of single quadruplets into two
doublets indicating the lifting of the valley degeneracy (Fig.~\ref{fig:One}b, d, e).
This identification of the multiplet character goes far beyond the results achieved
by chemical etching of monolayer graphene QDs \cite{Bischoff2015} or double-sided gating of bilayer graphene QDs \cite{Allen2012,Goossens2012,Mueller2014}.

%
%
\section*{Movable quantum dot}

Here, we explore the nanoscale variation of the charging sequence
in detail. We use a
heterostructure comprised of a SiO$_2$/graphite support, a hexagonal
boron nitride (hBN) substrate, and an active graphene layer on top which are assembled
by the dry stacking method \cite{Mayorov2011,Kretinin2014} (Fig.~\ref{fig:One}a). The atomic lattices of
graphene and hBN are collinearly aligned in order to create a hexagonal
superlattice with lattice constant $a=13.8$~nm originating from the
lattice mismatch of graphene and hBN.\cite{Woods2014}
Different stacking regions of the C atoms with respect to the B and N atoms (Fig. \ref{fig:One}c) naturally lead to a spatially varying
adhesion energy as well as to a spatially
varying sublattice symmetry breaking of graphene due to the inequivalent
binding sites. The resulting structure
has been extensively discussed in the literature \cite{Sachs2011,vanWijk2014,Chizhova2014,SanJose2014,Slotman2015,Jung2017}.
It is known that the most attractive interaction is in the AB areas (Fig.~\ref{fig:One}c) leading
to stretched central regions of graphene with AB stacking and closest
contact to the hBN. These areas are surrounded by compressed graphene ridges
of different stacking with larger separation to the
hBN \cite{vanWijk2014,Woods2014}. However, firm conclusions on the details of the superstructure are difficult to draw, because of the lack of knowledge of details
of the van-der-Waals interaction \cite{Ambrosetti2016}.

\begin{figure*}[h!]
	\centering
    \includegraphics[width = 0.5\textwidth]{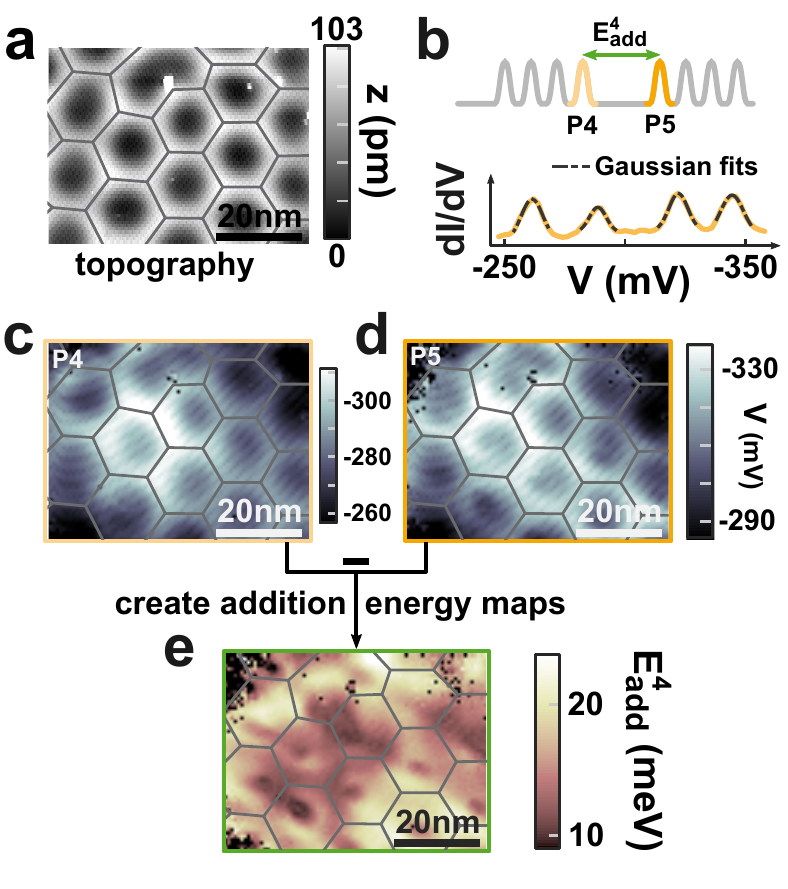}
    \caption{ \textbf{Addition energy maps from dI/dV spectra}.
	\textbf{(a)} STM image of graphene collinearly aligned to the h-BN substrate, $V = 400\mV$, $I = 300\pA$, $B=7\T$, $T=8$~K. The overlay of grey lines marks the supercell boundary deduced from the topography.
	\textbf{(b)} Top: sketched charging peak sequence with highlighted peaks P4 and P5 separated by addition energy $E_{\rm add}^4$. Bottom: typical $dI/dV$  curve (yellow line) with Gaussian fits (dashed lines) used to determine peak voltages $V_{{\rm P}n}$.
	\textbf{(c,d)} Maps of $V_{{\rm P}4}$ (\textbf{c}) and $V_{{\rm P}5}$ (\textbf{d}) of the area of \textbf{a} with identical grey lines overlaid, same parameters for measurement of the map of $dI/dV$ curves as in Fig.~\ref{fig:One}e. The slight shift of the observed patterns with respect to the grey lines is attributed to a small lateral shift ($\sim 2$~nm) of the tunneling atom with respect to the center of the QD \cite{Morgenstern2000}.
    \textbf{(e)} $E_{\rm add}^4$ map deduced by  $E_{\rm add}^4(\x)= e\cdot \eta  \left|V_{{\rm P}5}(\x) - V_{{\rm P}4}(\x) \right|$, same grey lines as in \textbf{a}, \textbf{c}, \textbf{d}.}
	\label{fig:Two_Exp}
\end{figure*}
%
The tip induced graphene QD can be
moved across the graphene superstructure by moving the STM tip \cite{Dombrowski1999}. This allows to tune the QD
properties, which we probe by tracking the position of the charging peaks within the superlattice. Therefore, we
employ spatially resolved $dI/dV$ spectra ($I$: tunneling current,
$V$: tip voltage). The resulting maps of charging energies can
be directly compared with the corresponding topographic maps recorded
simultaneously (Fig.~\ref{fig:Two_Exp}a). The charging peaks are
fitted by Gaussians (Fig.~\ref{fig:Two_Exp}b) for each QD center position
$\x$, rendering maps of the local variation of the voltage
$V_{{\rm P}n}(\x)$ of the $n^{\rm th}$ peak, ${\rm P}_n$ (Fig.~\ref{fig:Two_Exp}c, d). Typical variations between
the center and the boundary of the hexagonal supercell are
$\Delta V_{{\rm P}n}\approx 20$\,mV.  In order to relate this to an energy variation $\Delta E_n$ of
a particular QD level, we employ a capacitive model yielding $\Delta E_n = e \cdot \eta \cdot \Delta V_{{\rm P}n}$ with the
lever arm $\eta \simeq 0.5$ \cite{supplement} and electron charge $e$.
The $\Delta E_n$ variations are primarily caused by the spatially varying adhesion energy across the supercell, which indeed varies on
the 10 meV-scale according to extensive model calculations \cite{vanWijk2014}.
Figure~\ref{fig:Two_Exp}c and d additionally exhibit a long-range variation on the 50 nm scale (amplitude $\Delta
V_{{\rm P}n} \simeq 40$~mV) which we attribute to the uncontrolled,
long-range disorder potential of graphene on hBN with strength
of about 20~meV and correlation length of about 50~nm. Similar disorder potentials have been found previously \cite{XueJ2011, Decker2011}. Note that we carefully avoid lifting of the graphene
layer by the tip forces, i.e., we regularly record $I(z)$ curves ($z$:
tip-sample distance) verifying that the current remains below the threshold where a
slope change of $\ln{(I(z))}$ indicates
lifting \cite{Mashoff2010,Georgi2017}.

\section*{Tracking orbital, valley and spin splitting}
The group of the first four charging peaks, P1 to P4, is associated with the
quadruplet belonging to the first hole orbital of the QD. During the charging of these levels, the QD exhibits
a depth of about 100\meV and a width of about 50\nm as known from detailed Poisson calculations \cite{supplement,Freitag2016}. The confined wave functions are labeled  $\Psi_{\alpha,\tau,\sigma}$ with
orbital quantum number $\alpha=1$ for the first four peaks, valley quantum
number $\tau = \pm \frac 12$ and spin quantum number $\sigma=\pm\frac
12$. Analogously, the next four peaks, P5 to P8, belong to the
filling of the quadruplet $\Psi_{\alpha=2,\tau,\sigma}$. Subtracting
the voltage of the highest peak of the first quadruplet $V_{\rm{P}4}$ (Fig.~\ref{fig:Two_Exp}c) from
that of the lowest peak of the second, $V_{\rm{P}5}$ (Fig.~\ref{fig:Two_Exp}d), and multiplying by
$\eta$, yields the locally varying addition energy map
$E^4_{\rm add}(\x) = e \cdot \eta  |V_{\rm{P}5}(\x)-V_{\rm{P}4}(\x)|$ (Fig.~\ref{fig:Two_Exp}e). It consists of the charging energy $E_{\rm C}^4(\x)$ and the energy difference $E_{2,-\frac 12,-\frac 12}(\x) - E_{1,+\frac  12,+\frac 12}(\x)$.
The latter includes the valley splitting $E_{\alpha,+\frac 12,\sigma}(\x)-E_{\alpha,-\frac 12,\sigma}(\x)$ and the considerably
smaller Zeeman splitting $E_{\alpha,\tau,+\frac 12}(\x)-E_{\alpha,\tau,-\frac 12}(\x)= g\mu_{\rm B} B \approx 0.8$\meV ($g=2$:
gyromagnetic factor of graphene, $\mu_{\rm B}$: Bohr magneton). The dominant contribution comes from the orbital splitting $E_{2,\tau,\sigma}(\x)-E_{1,\tau,\sigma}(\x)$
as known from tight binding calculations \cite{Freitag2016}.
Since the wave function size does not change strongly as a function of $\x$ (see movie in supporting information), the spatial variation of $E_{\rm C}^4(\x)$
cannot explain the strong spatial variation of $E^{4}_{\rm add}(\x)$, which varies by a factor of two. Hence, $E_4^{\rm add} (\x)$ (Fig.~\ref{fig:Two_Exp}e) mostly maps out the orbital-energy
spacing between $\alpha=1$ and $\alpha=2$, as the quantum dot is moved
across the graphene superstructure. Periodic depressions in the center of the supercell reveal the influence of the superstructure on the orbital splitting, while the long-range
structure in Fig.~\ref{fig:Two_Exp}e (50 nm scale) is again attributed to the long-range potential disorder.

\begin{figure*}[h!]
		\centering
    \includegraphics[width = 1.0\textwidth]{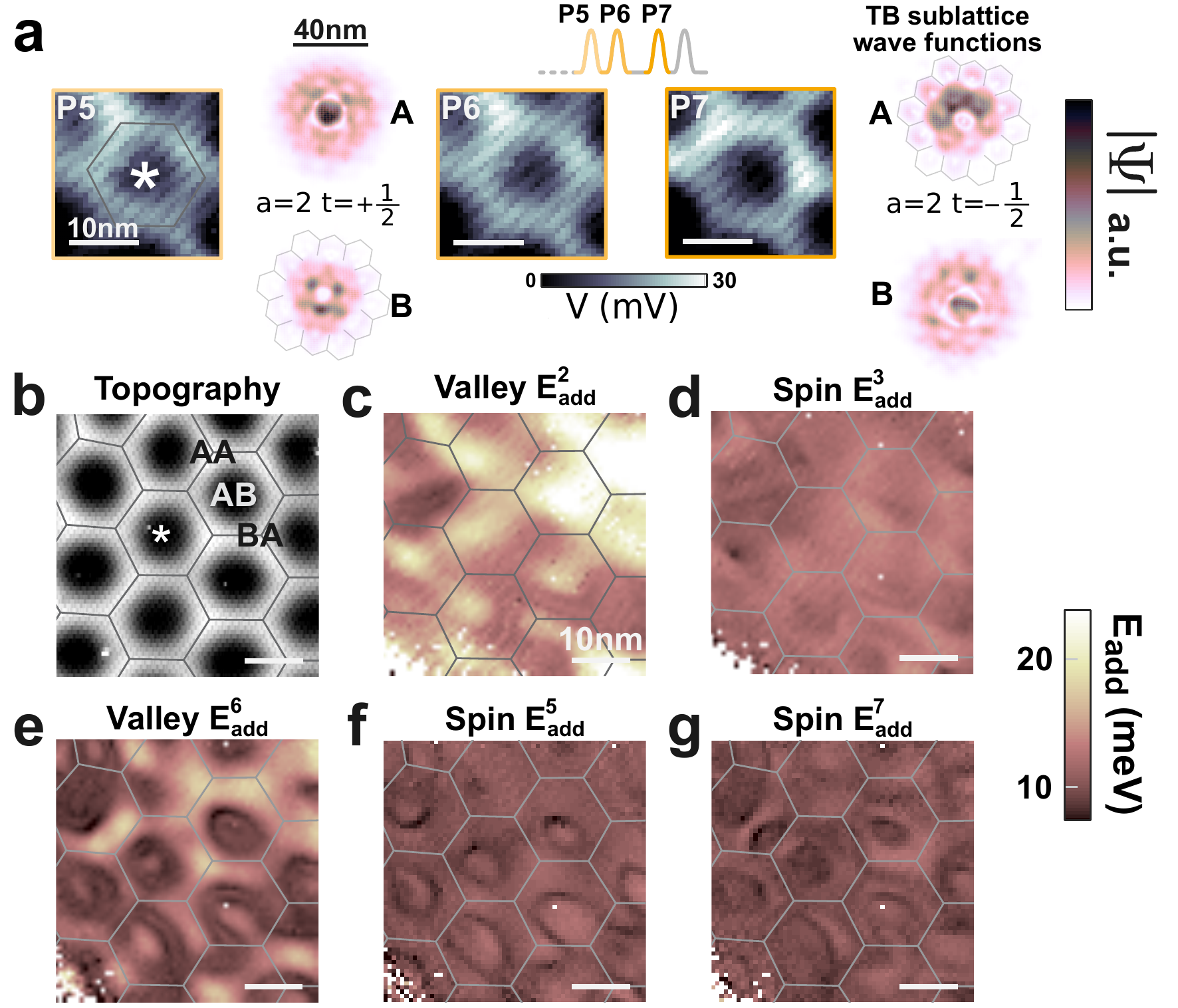}
    \caption{ \textbf{Addition energy maps for spin and valley gaps}.
      \textbf{(a)} $V_{{\rm P}n}(\x)$ displayed at identical contrast
      for $n=5,6,7$. The corresponding charging sequence is sketched
      on top. The diagonal stripes are caused by the atomic lattice of graphene via a moir{\'e} effect as outlined in supplementary section 6. Asterisk in P5 marks the identical position in
      \textbf{b}.  Also shown are the moduli of the wavefunctions for
      the second hole orbital, $\left|\Psi_{\alpha=2,\tau =+1/2}\right|$ and
      $\left|\Psi_{\alpha=2,\tau = -1/2}\right|$, decomposed into the two
      sublattice contributions, as marked by A and B, as calculated by our TB model. The quantum dot center is in the AB
      stacking region. Grey honeycombs mark the unit cells of the
      graphene superstructure. Note the different length scales of
      $|\Psi|$ maps and $V_{{\rm P}n}$ maps. \textbf{(b)} STM image of
      graphene on hBN including the area of \textbf{a}. Grey lines mark supercell boundaries. Different stacking areas (AA, AB, BA) are indicated, $V = 400\mV$, $I = 300\pA$. \textbf{(c)-(g)} $E_{\rm add}^n (\x)$ maps exhibiting
      identical contrast and belonging to valley and spin gaps as
      marked, same grey lines as in \textbf{b}. Length of all unlabeled scale bars in
      (\textbf{a}$-$\textbf{g}): 10\nm. Same parameters for $dI/dV$
      spectra as in Fig.~\ref{fig:One}e.  }
	\label{fig:Three}
\end{figure*}
%
For clarity, we focus now on the second hole orbital shell $\alpha=2$ (Fig.~\ref{fig:Three}), while we provide other $E_{\rm add}^n$ maps in the supplementary sections 4 and 5.
The local variation of the voltage peaks belonging to the $\alpha=2$
quadruplet allows
to map out valley and spin splittings in detail. The voltage maps, $V_{P6}$ and $V_{P7}$, differ on
length scales well below that of the supercell size ($\approx 10$~nm), and much smaller
than the size of the QD wave function (diameter $\approx 40$~nm, calculated by our TB approach) (Fig.~\ref{fig:Three}a).
The addition energy maps (Fig.~\ref{fig:Three}e$-$g) clearly display short-range supercell-periodic
variations on the length scale of $ 3$~nm. These variations appear as dark,
ring-like structures around the AB stacking region of the supercell in the valley addition energy map $E_{\rm add}^6$.
Similar, but slightly narrower rings  appear in the spin addition energy maps $E_{\rm add}^5$ and $E_{\rm add}^7$.

\section*{Analyzing the valley splitting maps}
We analyze these remarkably strong nanometer scale variations by
performing TB calculations \cite{Libisch2010, Chizhova2014}. The calculations feature three major ingredients: (i) the
sublattice-independent local on-site potential $V_0(\x)$ representing
the spatially varying adhesion energy, (ii) the sublattice
symmetry-breaking on-site potential $V_z(\x)$ caused by the
spatially varying stacking, and (iii) a locally varying hopping amplitude
$\gamma(\x)$ accounting for strain which also breaks sublattice symmetry
 \cite{CastroNeto2009,Georgi2017,Jung2017}. We use an average
distance between graphene and hBN of $3.3$~\AA, originating from DFT calculations employing the random phase
approximation \cite{Sachs2011} and consistent with cross sectional
electron microscopy data \cite{Haigh2012}. To obtain locally varying
tight-binding parameters, we first employ a continuum model of
graphene with known elastic constants \cite{SanJose2014} subject to
the potential landscape from the hBN \cite{vanWijk2014}. This
reproduces the corrugation of 70 pm and the strain variation of 2 \%, as visible in the STM data
(Fig.~\ref{fig:Two_Exp}a) \cite{Woods2014}. Based on the resulting membrane shape of the graphene layer, a molecular
dynamics simulation using isotropic Lenard-Jones potentials is employed to obtain
the atomically resolved strain, the variations in the local distance
between hBN and graphene, and the local stacking configuration
\cite{supplement}. Using these input parameters, we determine
$V_0(\x)$, $V_z(\x)$ and $\gamma(\x)$ from our own DFT calculations
\cite{supplement}. The potentials and hopping parameters provide, in
turn, the input to our third-nearest neighbor TB calculation of the QD states
\cite{Libisch2010, Chizhova2014,Freitag2016}. We emphasize that no
freely adjustable parameter enters our simulation. More details are described in supplementary sections 7-10.

%
\begin{figure*}[h!]
    \centering
    \includegraphics[width = \textwidth]{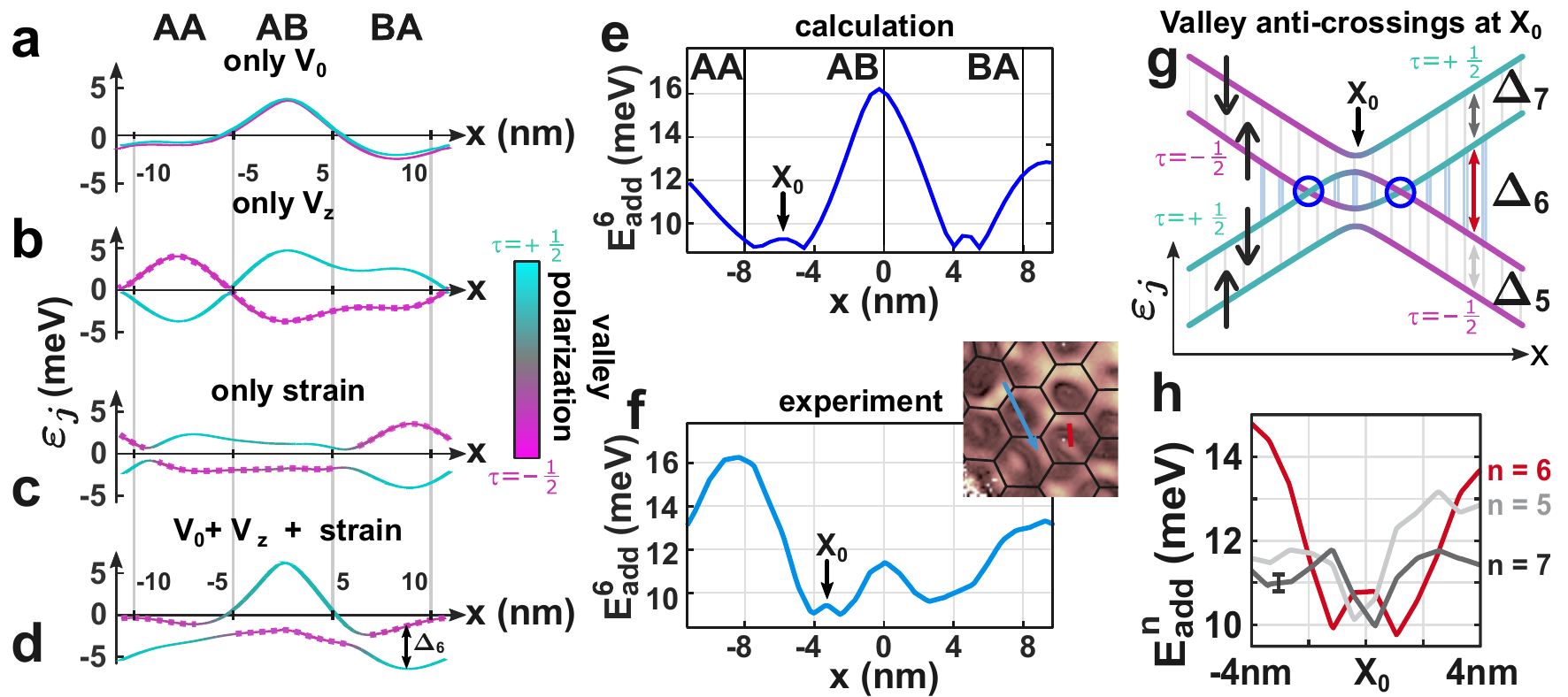}
    \caption{ \textbf{Valley crossing}.
      \textbf{(a)$-$(d)} TB energies ($\epsij$)
      of the two valley states of the second QD hole orbital ($\alpha =2$) as
      a function of the center position of the QD. Different stackings
      at this center along a high symmetry line of the superstructure
      are given on top. The valley polarization is
      color coded. Panel (\textbf{a}) considers only the sublattice
      independent potential $V_0(\x)$, (\textbf{b}) only the
      sublattice symmetry breaking on-site potential $V_z(\x)$,
      (\textbf{c}) only the varying hopping parameter $\gamma(\x)$ due
      to strain, and (\textbf{d}) the sum of all three
      contributions. The valley splitting $\Delta_6$ determining the spatial variation in
      \textbf{e} is indicated by a double arrow in \textbf{d},
      $B=7$~T. \textbf{(e)} Theoretical prediction for $E_{\rm add}^6$ based on
      \textbf{d} (supplementary section 8). \textbf{(f)} Experimental addition energy $E_{\rm
        add}^6$ along the arrow of the same color as in the inset
      (same $E_\mathrm{add}^6(\x)$ map as Fig.~\ref{fig:Three}d). The
      $x$-axis is aligned to the stackings marked in \textbf{e}. $X_0$
      indicates a feature attributed to the influence of spin
      splitting at the valley crossing. The origin in  \textbf{(a)$-$(f)} is chosen in the center of the AB region. \textbf{(g)} Schematic
      evolution of the state energies for a crossing of two
      valley states ($\tau = +1/2$: cyan, $\tau = -1/2$: magenta). A spatially
      constant spin splitting (levels marked by black spin arrows) is
      added. The resulting energy differences $\Delta_n$ are marked by
      double arrows. An anticrossing emerges at $X_0$ as deduced from \textbf{d}. Blue circles mark spin level crossings. \textbf{(h)} Experimental
      $E_{\rm add}^n(\x)$ along the red line in the inset of
      \textbf{f}, belonging to one preferential valley gap (red) and
      two spin gaps (grey). A typical error bar, resulting from the
      Gaussian fits of the $dI/dV$ peaks, is shown.}
		\label{fig:Four}
\end{figure*}

In agreement with the experiment, the calculated energies of the two
valley states of the second orbital feature a pronounced variation
with QD position (Fig.~\ref{fig:Four}a$-$d). To disentangle the
influence of strain and of the hBN substrate interaction, we analyze
the contributions due to $V_0(\x)$, $V_z(\x)$, and $\gamma(\x)$ separately. While $V_0$ (Fig.~\ref{fig:Four}a)
introduces local variations of the energy of the hole orbital $\alpha=2$
along the path AA$\leftrightarrow$AB$\leftrightarrow$BA,
it does not lift the fourfold valley and spin
degeneracy. $V_z(\x)$, by contrast, lifts the degeneracy
between the two valley states $\Psi_{2,+\frac 12,\sigma}$ and
$\Psi_{2,-\frac 12,\sigma}$ and even leads to an inversion of the energetic
order in the AA region of the superlattice, i.e., a change of sign
of $E_{2,+\frac 12,\sigma } - E_{2,-\frac
  12,\sigma}$ (Fig.~\ref{fig:Four}b). However, only when the
contribution of strain is accounted for through $\gamma(\x)$, which
inverts the sign of the valley splitting in the BA region (Fig.~\ref{fig:Four}c),
the correct overall level ordering with level inversion in the AB region,
as seen in our experiment, emerges (Fig.~\ref{fig:Four}d).

The addition energies in both, the TB model (Fig.~\ref{fig:Four}e) and the experiment (Fig.~\ref{fig:Four}f),
show the same variation of about $6$\,meV and the same order of maxima and minima along the displacement coordinate $x$.
Hence, we attribute the periodically appearing rings encircling the AB region (Fig.~\ref{fig:Three}e), which correspond to the bump at $X_0$ with adjacent minima in Fig.~\ref{fig:Four}e, 
as the positions of an inversion of valley ordering. Remaining quantitative differences between TB model and experiment (Fig.~\ref{fig:Four}e, f)
are attributed to disorder, most likely due to residual irregular strains caused by the non-perfect collinear alignment between graphene and hBN.
The resulting disorder is directly visible as irregularities in the unit cell of the superstructure (Fig.~\ref{fig:Two_Exp}a, Fig.~\ref{fig:Three}b) and also explains the irregular distortions of the rings around the AB region.

The assignment of the rings around the AB region to valley inversions is corroborated by the
appearance of the small bump in the ring minimum, marked $X_0$ in Fig.~\ref{fig:Four}e$-$h.
It is found in theory and experiment
with a height of less than 1 meV. The
theoretical level diagram (Fig.~\ref{fig:Four}g) provides a simple
explanation: the bump is the result of the additional spin splitting during the passage through the
crossing of valley levels. At $X_0$, $E^6_{\rm add}$ consists of $E_{\rm C}^6$ and the spin
splitting  $\left|E_{2,\tau,\frac 12} -
E_{2,\tau,-\frac 12}\right| \approx 800\;{\rm \mu}$eV reduced by anti-crossing contributions.
In contrast, the two spatially offset crossings of valley states with different spins (blue circles in Fig.~\ref{fig:Four}g) 
feature only $E_{\rm C}^6$, resulting in the minima around the bump.
Figure~\ref{fig:Four}g also explains the rings in the spin splitting maps (Fig.~\ref{fig:Three}f, g), which are simply the reduced $\Delta_5$ and $\Delta_7$
at $X_0$. The spatial alignment of the bump in $\Delta_6$ and the minima in $\Delta_{5,7}$ are nicely corroborated by the experiment (Fig.~\ref{fig:Four}h).

While we have focused here on the valley splitting of the second hole state, similar ring-like structures encircling the AB area are also
found for the third hole orbital $\alpha=3$ with tunability of the valley
crossing up to 15~meV (Fig.~S2) \cite{supplement}.
In contrast, the first hole orbital $\alpha=1$ (Fig.~\ref{fig:Three}c, e) exhibits a valley tunability of about 7 meV without inversion of the valley ordering.
On the electron side, the additional charging of defects within the h-BN
\cite{Wong2015} complicates the analysis \cite{Morgenstern2017}, but
some ring-like structures indicating valley inversion can also be
spotted \cite{supplement}. Data recorded
with another microtip at two different $B$ fields exhibit very similar features (Fig.~S3). Moreover, the energy range of valley tunability remains independent of $B$, corroborating that the valley tuning is caused by the interaction with the substrate and not by the $B$ field. For example, the strength of the exchange enhancement would vary with $B$ \cite{supplement}. In addition, it turns out to be one order of magnitude too weak to explain the experimentally observed valley tuning (supplementary section 12).

A simple estimate clarifies the
resulting strength of the valley splitting of about 10 meV. The sublattice breaking interactions itself ($V_z(\x)$, $\gamma(\x)$) spatially vary by about
100 meV as deduced from our DFT calculations \cite{supplement}. Hence, shifting about 10\% of the hole
density of a state ($\propto \left|\Psi\right|^2$) from the unfavorable AB to the favorable AA region is sufficient to account for
variations of the valley splitting of about 10 meV.
Indeed, our detailed TB calculations find that the $\alpha=2$ wave function covers about ten unit cells (Fig.~\ref{fig:Three}a)
and adjusts mainly its distribution within the central unit cell
due to the changing potential landscape (supplemental movie).

\section*{Conclusions and outlook}
The revealed tunability of a valley splitting by up to $15$~meV surpasses the highest
reported values of valley tuning for other potentially nuclear spin
free host materials (Si/SiO$_2$, $500\,\mathrm{\upmu
  eV}$) \cite{YangC2013} by more than one order of magnitude. Hence, it might
be exploited at temperatures up to 4~K.
Most intriguingly, the crossings of valley and spin levels as depicted in Fig. \ref{fig:Four}g can be used to
initialize superposition states of spin and valley degrees of freedom \cite{PieroLadriere2008,Petta2005}.
This could be the starting point to determine the coherence \cite{Elzerman2004} of both types of states in graphene for the first time.
The required interaction of the levels rendering the depicted crossings into anti-crossings is naturally provided
by the spatially varying sublattice potential coupling opposite valley states (Fig.~\ref{fig:Four}d). We note in passing that
the breaking of the valley degeneracy is also the central requirement for exchange-based spin qubits, which could
provide an all electrical spin qubit operation in graphene \cite{Trauzettel2007}.
A possible device setup for these purposes could employ side gates for moving gate-based QDs and, hence, for
providing the valley tuning. Edge states, belonging to each LL, can provide tunable source and drain contacts (supplementary section 13).

Finally, we emphasize that the approach of designed van-der-Waals heterostructures
\cite{Geim2013,Woods2014,Novoselov2016} for a versatile tuning of electronic degrees of
freedom might be extended to physical spin schemes by using an atomically varying spin orbit interaction as present, e.g.,
for graphene on WSe$_2$ \cite{Wang2016}.
\par

\subsection*{Acknowledgements}
The authors appreciate helpful discussions with C.~Stampfer, H.~Bluhm, R.~Bindel,
M.~Liebmann, and K.~Fl\"ohr as well as the assisting during the measurements by A.~Georgi.
NMF, PNI and MM acknowledge support from the European Union
Seventh Framework Programme under Grant Agreement No. 696656 (Graphene Flagship) and the
German Science foundation (Li 1050-2/2 through SPP-1459),
LAC, JB and FL from the Austrian Fonds zur F\"orderung der
wissenschaftlichen Forschung (FWF) through the SFB 041-ViCom
and doctoral college Solids4Fun (W1243).
TB calculations were performed on the Vienna Scientific Cluster.
RVG, AKG and KSN also acknowledge support from EPSRC
(Towards Engineering Grand Challenges and Fellowship programs),
the Royal Society, US Army Research Office, US Navy Research
Office, US Airforce Research Office. KSN is also grateful to
ERC for support via Synergy grant Hetero2D. AKG was supported
by Lloyd’s Register Foundation. PNI acknowledges the support from the Hungarian Academy of Sciences Lend\"ulet grant no. LP2017-9/2017.

\subsection*{Author contributions}
N.M.F. carried out the STM measurements with assistance of P.N.I. and C.H. and evaluated the experimental data under supervision of P.N.I. and M.M.;  P.N.I. has performed the strain calculations, while T.R., F.L., and L.A.C. have contributed DFT and TB calculations; C.R.W., Y.C., R.V.G., A.K.G. and K.S.N. provided the sample; M.M. conceived and coordinated the project partly together with N.M.F., P.N.I. and F.L.; the comparison between theory and experiment has been conducted by N.M.F., M.M., F.L., and P.N.I.; M.M., N.M.F., P.N.I. and F.L. wrote the manuscript with contributions from all authors.

\subsection*{Methods}
	The sample was prepared by exfoliating graphite flakes on a SiO$_2$ substrate, followed by two consecutive dry transfers\cite{Mayorov2011,Kretinin2014} of $30\nm$ thick hexagonal hBN and monolayer graphene, respectively.
	During the graphene transfer, we took care to minimize the angular misalignment between the graphene lattice and the hBN lattice. Remaining small misalignments in the $0.1^\circ$ regime cannot be excluded \cite{Woods2014}.
	Moreover, a few small bubbles between the graphene and the h-BN apppear after transfer (see chapter S7 of Ref. \cite{Georgi2017}). Both of these effects lead to mechanical stresses which perturb the graphene/BN superlattice in period or shape \cite{Jiang2017} (Fig. 2a and 3b of the main text).
	The	graphene flake overlaps the hBN completely.
	This avoids	insulating areas, which would be hazardous to the STM tip, but does not allow for back-gate operation.
	Finally, electrical Cr/Au contacts ($2\nm$/$100\nm$) are evaporated onto the
	large bottom graphite flake via a shadow mask. Optical images of the device structure are available in the supplement of a previous publication \cite{Freitag2016}.

	STM and STS measurements are performed in a home-built ultra-high-vacuum STM chamber operating at temperature $T=8$ K and in magnetic fields up to $B=7$ T perpendicular to the surface.\cite{Mashoff2009} Tungsten tips are prepared by etching of W wires, which are subsequently controlled with an optical microscope. The microtips are transferred into the STM within the UHV chamber, where they are reshaped by controlled indentation into the Au(111) surface of a Au bead.\cite{Voigtlaender2008}
	Thereby, they form a Au apex of a few $10\nm$ in length as crosschecked by electron microscopy.
	We characterize the tips in-situ by mapping the topographic and spectroscopic features of the Au(111) surface prior to exchanging the Au crystal by the graphene sample.
	Scanning tunneling microscopy (STM) images are recorded in constant current mode at tunneling current $I$ and tip voltage $V$.
	Differential conductance  curves $dI/dV(V)$ are recorded by lock-in detection using a modulation voltage with amplitude $V_\mathrm{mod} = 2-5\mV_\mathrm{rms}$ and frequency \mbox{$f_\mathrm{mod} = 1223\,\mathrm{Hz}$}. After stabilizing the tip-sample distance at stabilization voltage $V_\mathrm{stab}$ and stabilization current $I_\mathrm{stab}$,
	the feedback loop is opened for the $dI/dV(V)$ recording. During the recording, the tip-sample distance is changed at a rate of  $50$~pm/V, approaching the sample by \mbox{$0.5$\,\AA} while sweeping $V$ from $1\V$ to $0\V$ and retracting it by the same distance while continuing to $-1\V$. This compensates for the changing height of the tunneling barrier as a function of $V$.\cite{Feenstra1994}
	The resulting change in tip-sample capacitance is below $2.5\,$\%.\cite{Freitag2016} It is, thus, neglected, since much smaller than other capacitance uncertainties (see below).
	Additionally, we normalize the $dI/dV$ data according to $(dI/dV(V))/I(V_{\rm stab})$ with $I(V_{\rm stab})$ being the firstly detected current after opening the feedback loop. This compensates for the influence of vibrations during the stabilization process.\\
	We focus on the first two orbital states for confined holes, originating from LL$_{-1}$ (see below), as they capture the essential features.
	On the electron side, charging of randomly distributed defects in the hBN \cite{Wong2015} impedes an unambiguous analysis of the QD charging patterns (see Fig.~S2 of the supplement) \cite{Freitag2016}.

\subsubsection*{Data availability}
The data that support the plots within this paper and other findings of this study are available from the corresponding author upon reasonable request.

\subsection*{Additional information}
Supplementary information is available in the online version of the paper. Reprints and permission information is available online at www.nature.com/reprints. Correspondence and requests for materials should be addressed to M.M.



\end{document}